\documentclass[12pt]{iopart}

\usepackage{psfig}

\def\gsim{\;\rlap{\lower 2.5pt
\hbox{$\sim$}}\raise 1.5pt\hbox{$>$}\;}
\def\lsim{\;\rlap{\lower 2.5pt
   \hbox{$\sim$}}\raise 1.5pt\hbox{$<$}\;}

\begin{document}

\title[Probing Distant Massive Black Holes with LISA]{Probing Distant
Massive Black Holes with LISA}

\author{Kristen Menou\footnote[1]{Celerity Foundation Fellow}
}

\address{Department of Astronomy, P.O. Box 3818, University of
Virginia, Charlottesville, VA 22903, USA}

\begin{abstract}
Idealized models are used to illustrate the potential of the Laser
Interferometer Space Antenna (LISA) as a probe of the largely unknown
population of cosmologically-distant Massive Black Holes (MBHs) and as
a tool to measure their masses with unprecedented accuracy.  The
models suggest that LISA will most efficiently probe a MBH population
of lower mass than the one found in bright quasars and nearby galactic
nuclei. The mass spectrum of these MBHs could constrain formation
scenarios for high-redshift, low-mass galaxies.
\end{abstract}


\submitto{\CQG}


\section{Introduction}

One of the major goals of the LISA experiment is the detection of the
gravitational wave signal from massive black hole (MBH) coalescences
at cosmological distances (see {\tt http://lisa.jpl.nasa.gov/}).
Despite remarkable progress in recent years, the characteristics of
the population of distant MBHs remains largely unknown. In the future,
LISA should open the ``gravitational window'' and offer us an usually
sharp view of this population. Idealized models of the MBH population
and its evolution with cosmic time are valuable in that they help
define the various ingredients important for the interpretation of the
future LISA data. In this contribution, I describe a class of such
models and highlight some of their most important characteristics.

\section{The population of massive black holes}

Recent years have seen tremendous progress in the characterization of
MBHs residing at the centers of nearby galactic nuclei (Kormendy \&
Richstone 1995). Dynamical evidence for the presence of MBHs in
galactic spheroidal components has been found in nearly all studied
local massive galaxies (Magorrian et al.  1998). A tight correlation
between the inferred BH mass and the spheroid stellar velocity
dispersion has also been established (Ferrarese \& Merritt 2000;
Gebhardt et al. 2000; Tremaine et al. 2002).

Much less is known about the cosmologically-distant population of
MBHs, however. Studies of the optical quasar luminosity function show
that, at redshifts $z \sim 2-3$ (corresponding to the peak of quasar
activity), quasars and associated MBHs are present in about $0.1\%$ of
bright galaxies at that epoch (Richstone et al. 1998). Recently, more
direct X-ray studies with {\it Chandra} revealed that $\sim 10\%$ of
all bulge-dominated, optically-luminous galaxies at $z \lsim 2-3$ show
central hard X-ray activity, which is presumably associated with
accretion onto a MBH (Mushotzky et al. 2000, Barger et al. 2001).

These useful constraints, at $z \lsim 3$, are consistent with the idea
that MBHs may be even rarer at higher redshifts. Several formation
scenarios for MBHs postulate that it is indeed the case (see, e.g.,
Eisenstein \& Loeb 1995; Volonteri et al. 2002).  A rare population of
MBHs would result in reduced event rates for LISA: of all the
successive galaxy mergers occurring in standard hierarchical structure
formation scenarios, only those involving a pair of MBHs can
potentially lead to a merger event detectable by LISA. As we shall see
below, this property can be inverted, in principle, so that the rate
of events detected by LISA would become a sensitive probe of the
population of distant MBHs.

\begin{figure}
\begin{center}
\hspace{-3cm}
\begin{minipage}[t]{0.45\hsize}
\begin{displaymath}
\psfig{figure=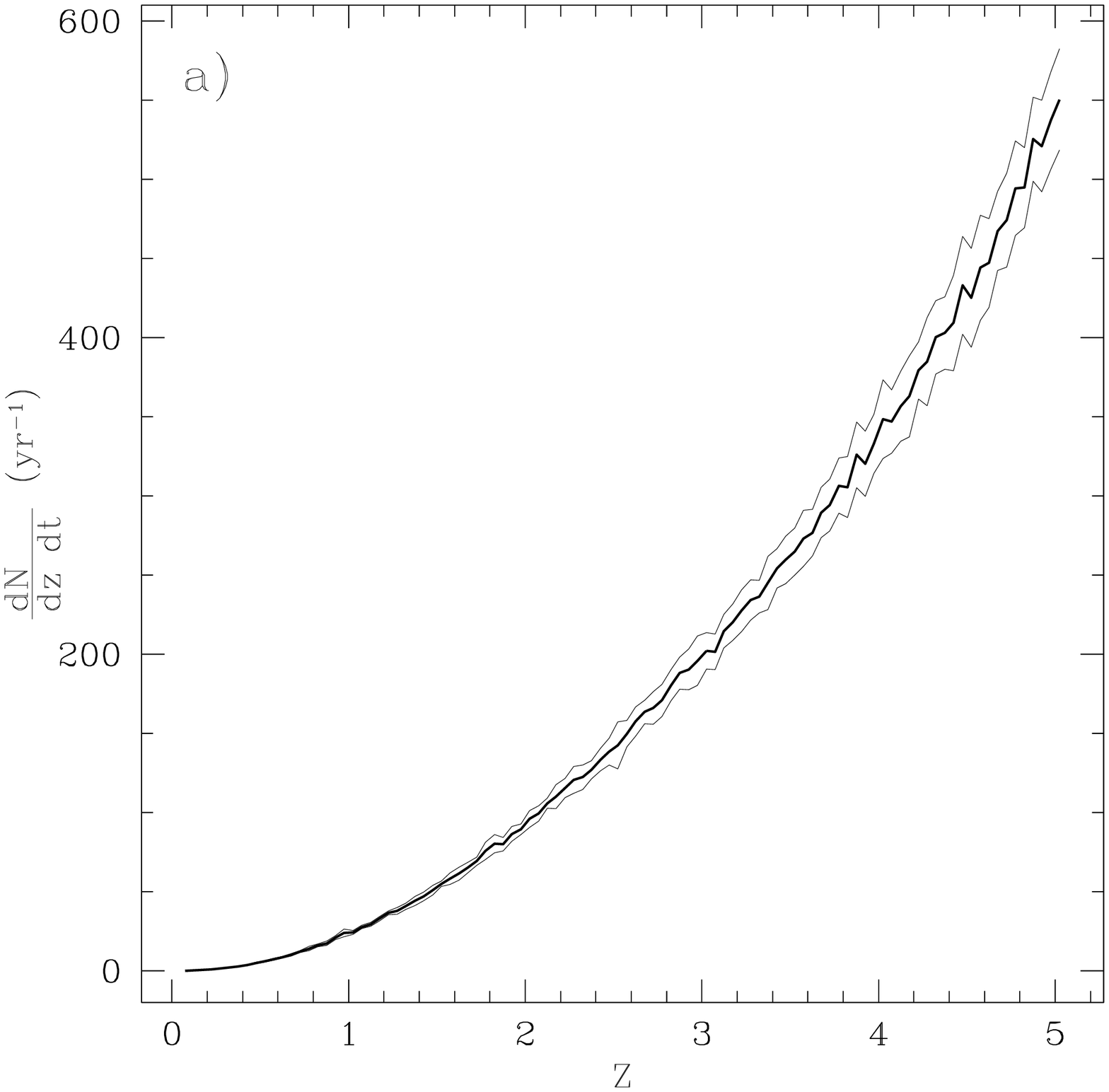,height=2.3in}
\end{displaymath}
\end{minipage}
\begin{minipage}[t]{0.45\hsize}
\begin{displaymath}
\psfig{figure=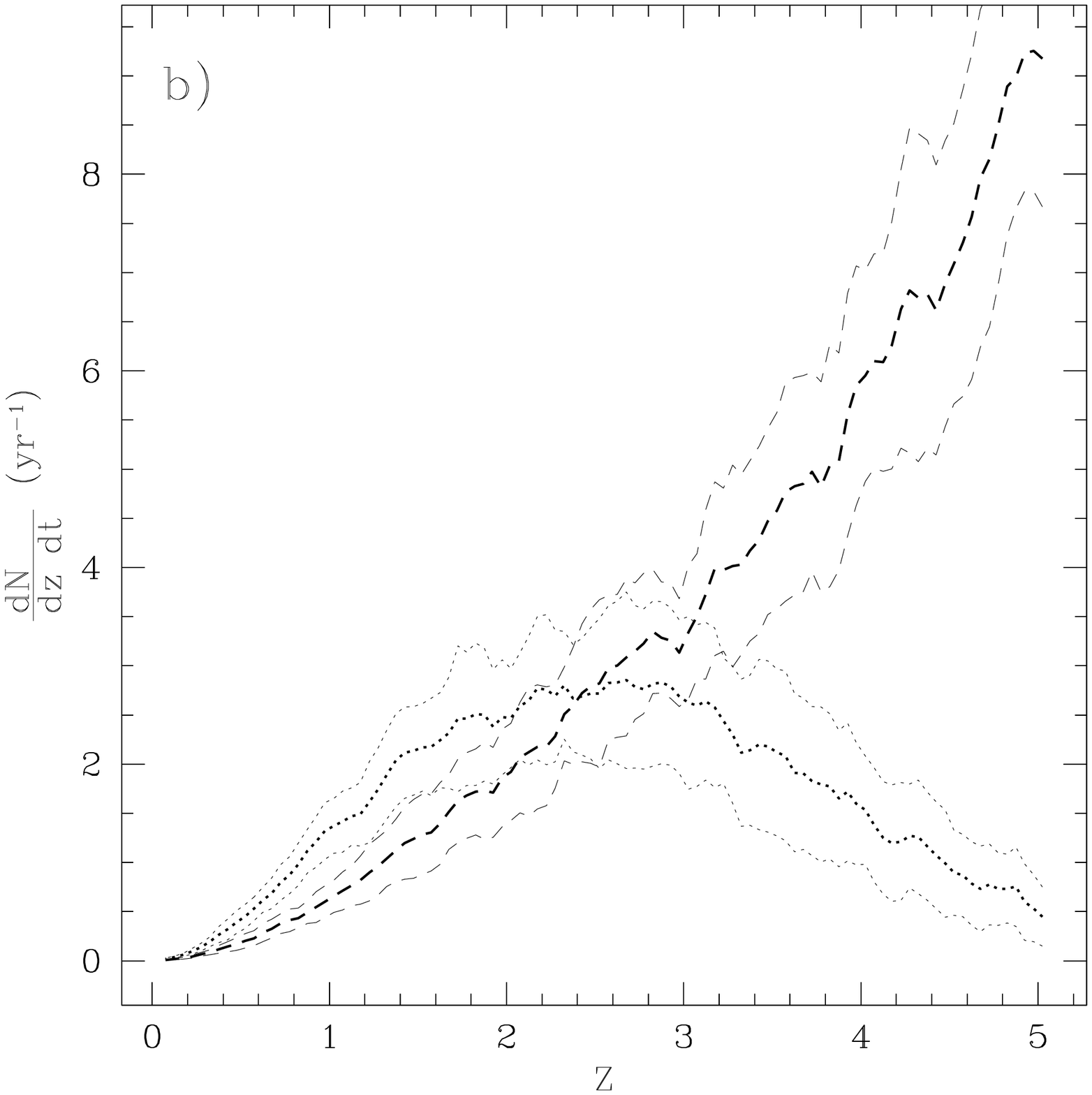,height=2.3in}
\end{displaymath}
\end{minipage}
\end{center}
\caption{\label{fig:one} Event rates (per year, per unit redshift) of
MBH mergers in models with BHs in all (100\%, a) or only 3\% (b) of
potential host galaxies at $z=5$. Very efficient MBH binary
coalescence is assumed.  In (b), two models are shown depending on
whether rare MBHs preferentially populate massive $z=5$ galaxies
(dashed) or populate them randomly (dotted).  These rates are likely
overestimated, as explained in the text. [From Menou et al. 2001.]}
\end{figure}

\begin{figure}
\begin{center}
\hspace{-3cm}
\begin{minipage}[t]{0.45\hsize}
\begin{displaymath}
\psfig{figure=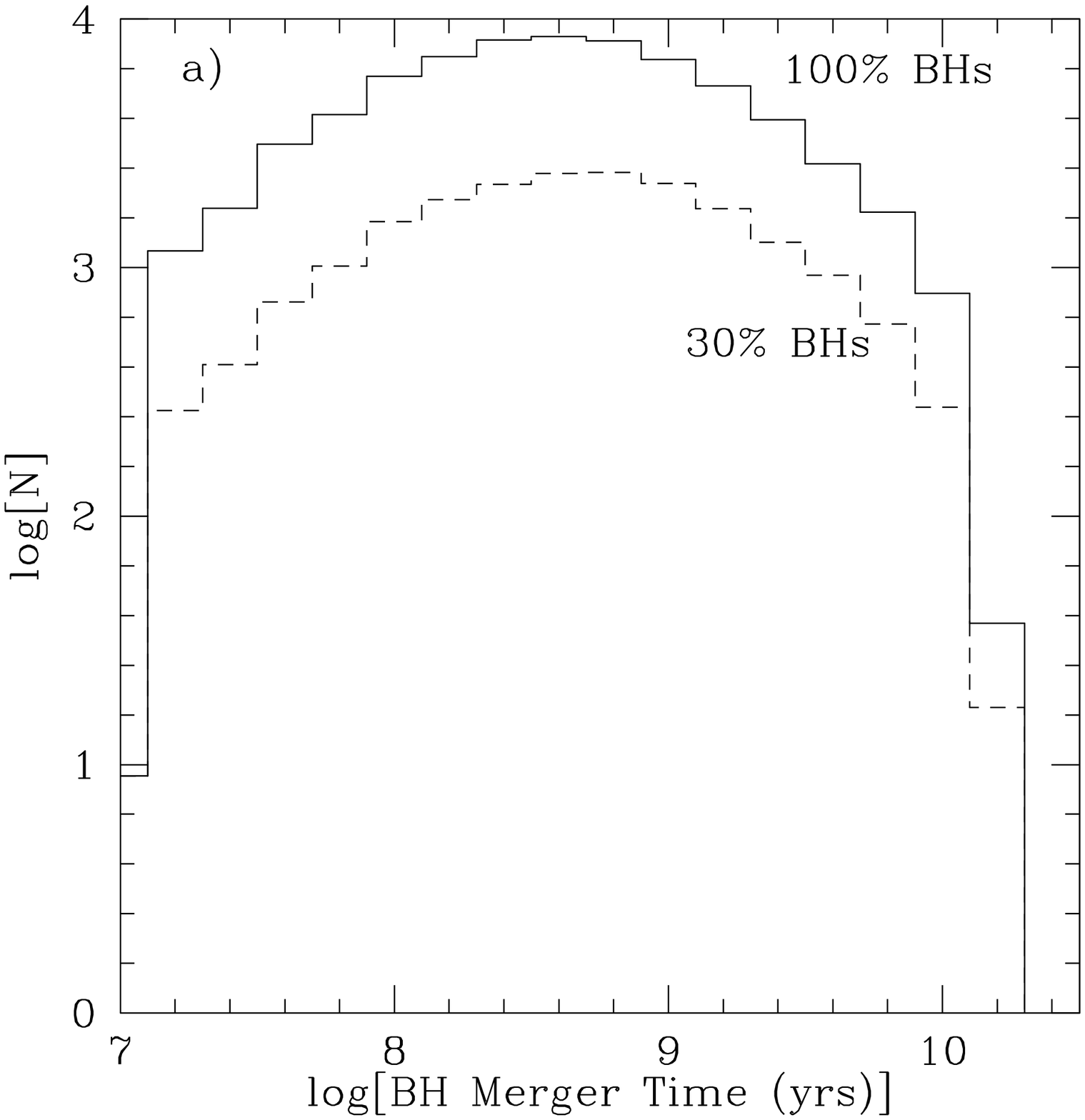,height=2.3in}
\end{displaymath}
\end{minipage}
\begin{minipage}[t]{0.45\hsize}
\begin{displaymath}
\psfig{figure=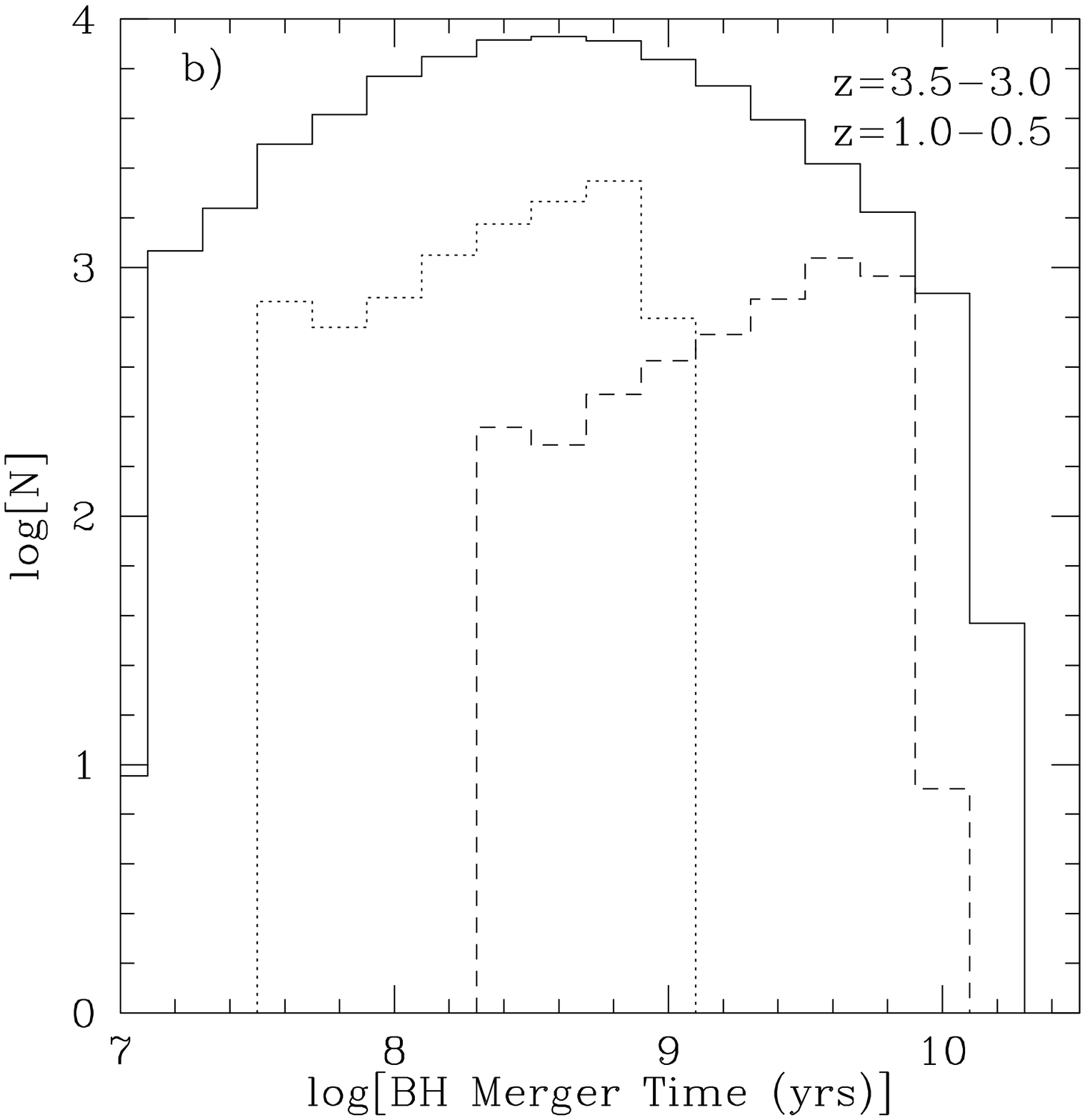,height=2.3in}
\end{displaymath}
\end{minipage}
\end{center}
\caption{\label{fig:two}(a) Distribution of time between successive
mergers of galaxies containing MBHs, integrated from $z=5$ to $0$, in
models with BHs in 100\% (solid) and only the 30\% most massive
(dashed) potential host galaxies at $z=5$. (b) Shows the contribution
to the total distribution (solid) of mergers in the redshift range
$z=3.5$--$3$ (dotted) and $z=1$--$0.5$ (dashed) for the model with BHs
in 100\% of potential host galaxies at $z=5$. Normalization is for a
fixed comoving volume of $\sim 1.7 \times 10^4$~Mpc$^3$.}
\end{figure}

\section{Event rate models and uncertainties}

The consequences of a rare population of MBHs at high redshifts have
been further investigated by Menou et al. (2001; see also Volonteri et
al. 2002). A standard "merger tree" was used to describe the merger
history of dark matter halos and associated galaxies in the
$\Lambda$CDM concordance cosmology ($\Omega_0 = 0.3$, $\Omega_b
=0.04$, $\Omega_{\Lambda} =0.7$, $h_{100}=0.65$). Given the local
constraint that nearly all galaxies more massive than $\sim 10^{11}
M_\odot$ (baryon + dark matter) must harbor a central MBH, as shown by
dynamical studies for a large enough sample of nearby galaxies with
masses $\gsim 10^{11} M_\odot$ (Magorrian et al. 1998), models with
rare MBHs showed that at least a few \% of all galaxies susceptible to
harbor a MBH must do so at $z=5$ (the tree's initial redshift). It was
found that this local constraint on the extent of the MBH population
is more stringent for the models than the other two based on optical
quasar and X-ray studies (simply because relatively few galactic
mergers occur at $z < 2$--$3$). Additional details can be found in
Menou et al. (2001).

Specifically, Menou et al. (2001) explored three models in which MBHs
populate 100\%, the 3\% most massive or a random (mass-independent)
3\% of all potential host galaxies at $z=5$.  The MBH merger event
rates corresponding to these three models are shown in
Fig.~\ref{fig:one}a and \ref{fig:one}b. The model with a maximal
population of MBHs (Fig.~\ref{fig:one}a) predicts rates about two
orders of magnitude larger than the models with a MBH population about
as rare as allowed by the local constraint (Fig.~\ref{fig:one}b).

Although these merger rates are very encouraging for LISA, they are
likely overestimated. First, LISA will be sensitive to a finite range
of BH masses, so that some of the events counted in Fig.~\ref{fig:one}
will be missed. Second, the orbital dynamics of two MBHs following the
merger of their host galaxies is rather uncertain. The inefficiency of
dynamical friction or subsequent nuclear stellar ejections at bringing
MBHs together (Begelman et al. 1980; Quinlan \& Hernquist 1997;
Milosavljevic \& Merritt 2001; Yu 2002) may imply that some of the
galactic mergers counted in Fig.~\ref{fig:one} are not actually
followed by prompt mergers of the resident MBHs.

Fig.~\ref{fig:two}a compares the time between successive mergers of
galaxies containing MBHs in models with BHs populating 100\% (solid
line) or only the 30\% most massive (dashed line) potential host
galaxies at $z=5$. In both models, a large number of successive
galactic mergers occur on timescales $\lsim 10^9$~yrs, leaving only
that much time for a MBH binary formed from a previous galactic merger
to coalesce. If a pre-existing MBH binary is unable to merge before a
third MBH makes its way to the galactic center, a three-body
interaction would result, leading typically to the slingshot ejection
of the least massive BH (Saslaw et al. 1964).

It is also worth noting that the present models assume that all the
galaxies described by the merger tree (with virial temperatures in
excess of $10^4$~K; see below) can potentially harbor a MBH.  There is
circumstantial evidence, however, that bulge-less galaxies may not
harbor such MBHs (Gebhardt et al. 2001; Merritt et al. 2001; but see
Filippenko \& Sargent 1989 for a possible counter-example). Accounting
for this would further reduce the event rates, in proportion to the
size of bulge-less galaxies (which may be significant at the
low-luminosity end; Bingelli et al. 1988).

\begin{figure}
\begin{center}
\hspace{-3cm}
\begin{minipage}[t]{0.45\hsize}
\begin{displaymath}
\psfig{figure=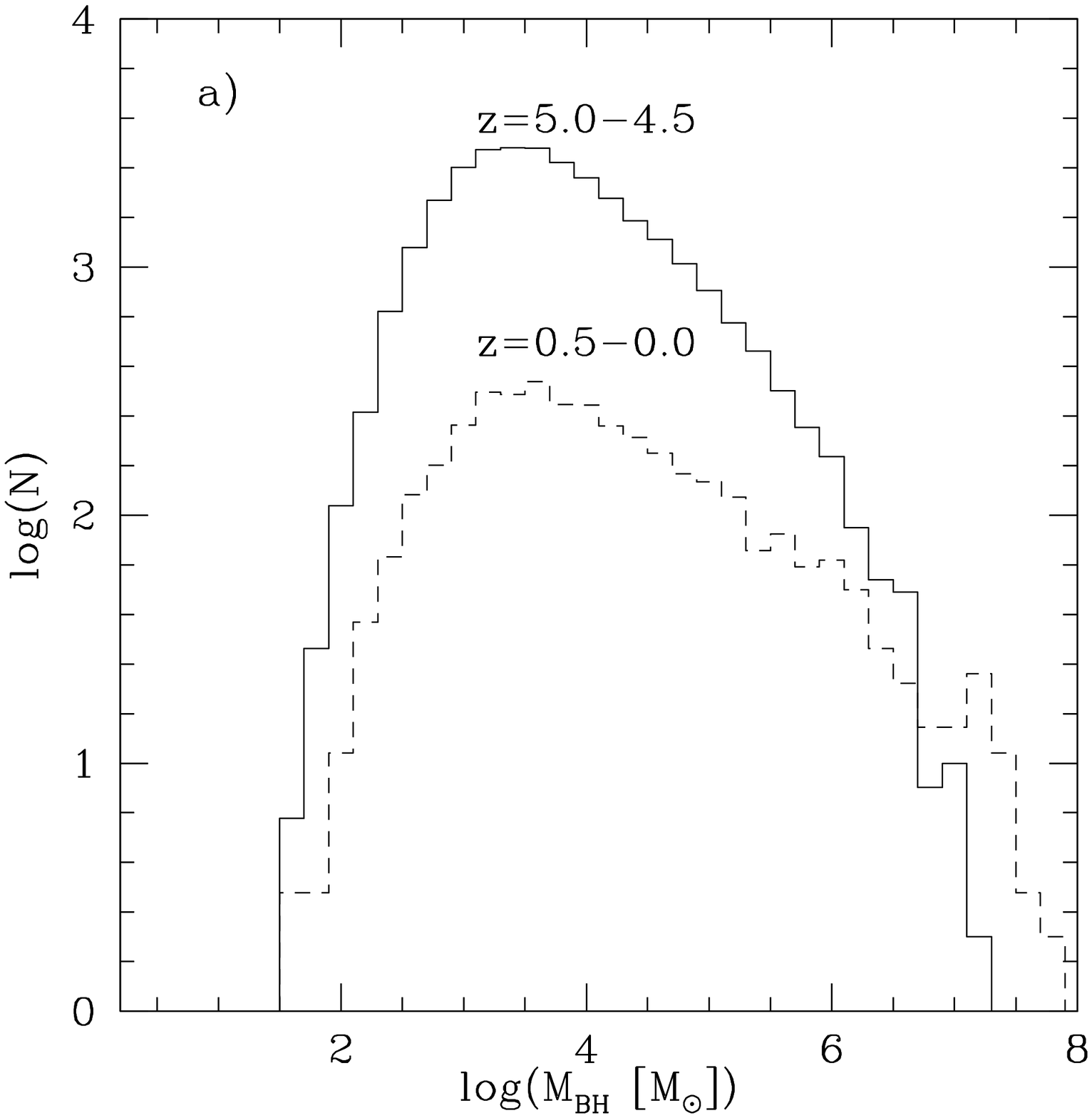,height=2.3in}
\end{displaymath}
\end{minipage}
\begin{minipage}[t]{0.45\hsize}
\begin{displaymath}
\psfig{figure=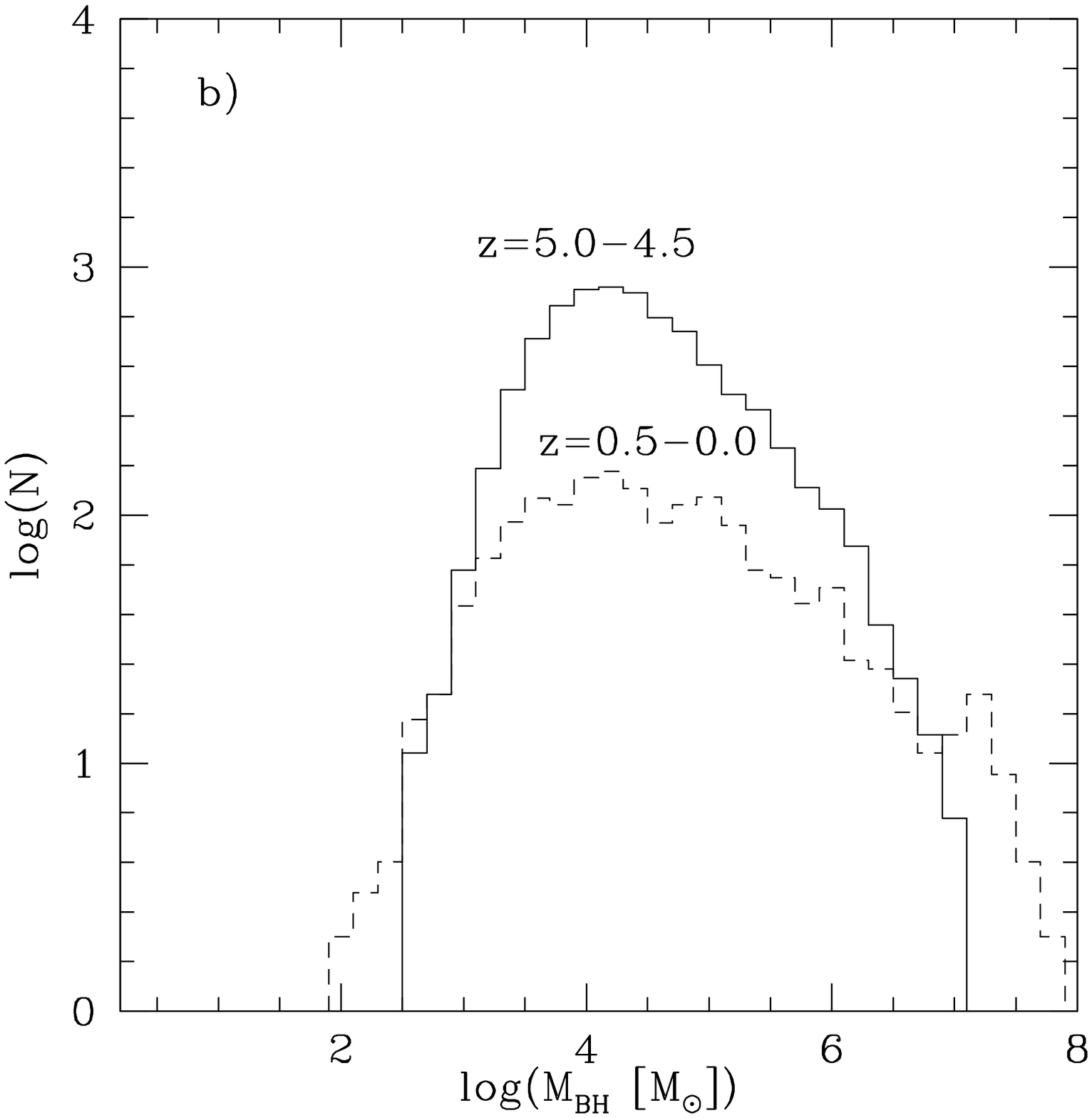,height=2.3in}
\end{displaymath}
\end{minipage}
\end{center}
\caption{\label{fig:three}Mass distributions of merging BHs in the
redshift range $z=5$--$4.5$ (solid) and $z=0.5$--$0$ (dashed), for the
models with BHs in 100\% (a) and only the 30\% most massive (b)
potential host galaxies at $z=5$. Normalization is for a fixed
comoving volume of $\sim 1.7 \times 10^4$~Mpc$^3$.}
\end{figure}

\section{Precision mass measurements}

LISA will not only be able to detect MBH coalescences, but it will
also constrain, and in some cases measure, the masses of the MBHs
involved. To illustrate the potential of LISA for precision mass
measurements, we use the same two models as described in the previous
section (with BHs in 100\% and only the 30\% most massive potential
host galaxies at $z=5$, respectively). At every redshift step in the
merger tree, the MBHs are forced to follow the mass -- velocity
dispersion relation with scatter (Tremaine et al. 2002):
\begin{equation}
M_{\rm BH}=\left( 1.35 \pm 0.2 \right) \times 10^8 M_\odot \left(
\frac{\sigma_e}{200~{\rm km s^{-1}}} \right)^{4.02 \pm 0.32},
\label{eq:one}
\end{equation}
where $\sigma_e$ is the stellar velocity dispersion of the spheroidal
component, at the half-light (effective) radius. It is related to
$\sigma_{\rm DM}$, the dark matter halo velocity dispersion, via the
relation $\sigma_e = \sigma_{\rm DM}/\sqrt(3/2)$, which is derived
through the Jeans equation for isotropic, spherical systems, with the
extra assumptions of an isothermal density profile ($\rho \propto
r^{-2}$) for the dark matter and a typical DeVaucouleurs density
profile ($\rho \propto r^{-3}$) for the stellar spheroidal
component. The dark matter halo velocity dispersion is obtained from
the virial theorem and the assumption that halos have a universal
density (evolving as $(1+z)^3$): $\sigma_{\rm DM} \propto M_{\rm
halo}^{1/3} (1+z)^{1/6}$. It is assumed that every single galaxy
described by the merger tree potentially harbors a MBH and no attempt
is made to separate a bulge-less galactic population potentially
unable to harbor such MBHs.

\begin{figure}
\begin{center}
\hspace{-3cm}
\begin{minipage}[t]{0.45\hsize}
\begin{displaymath}
\psfig{figure=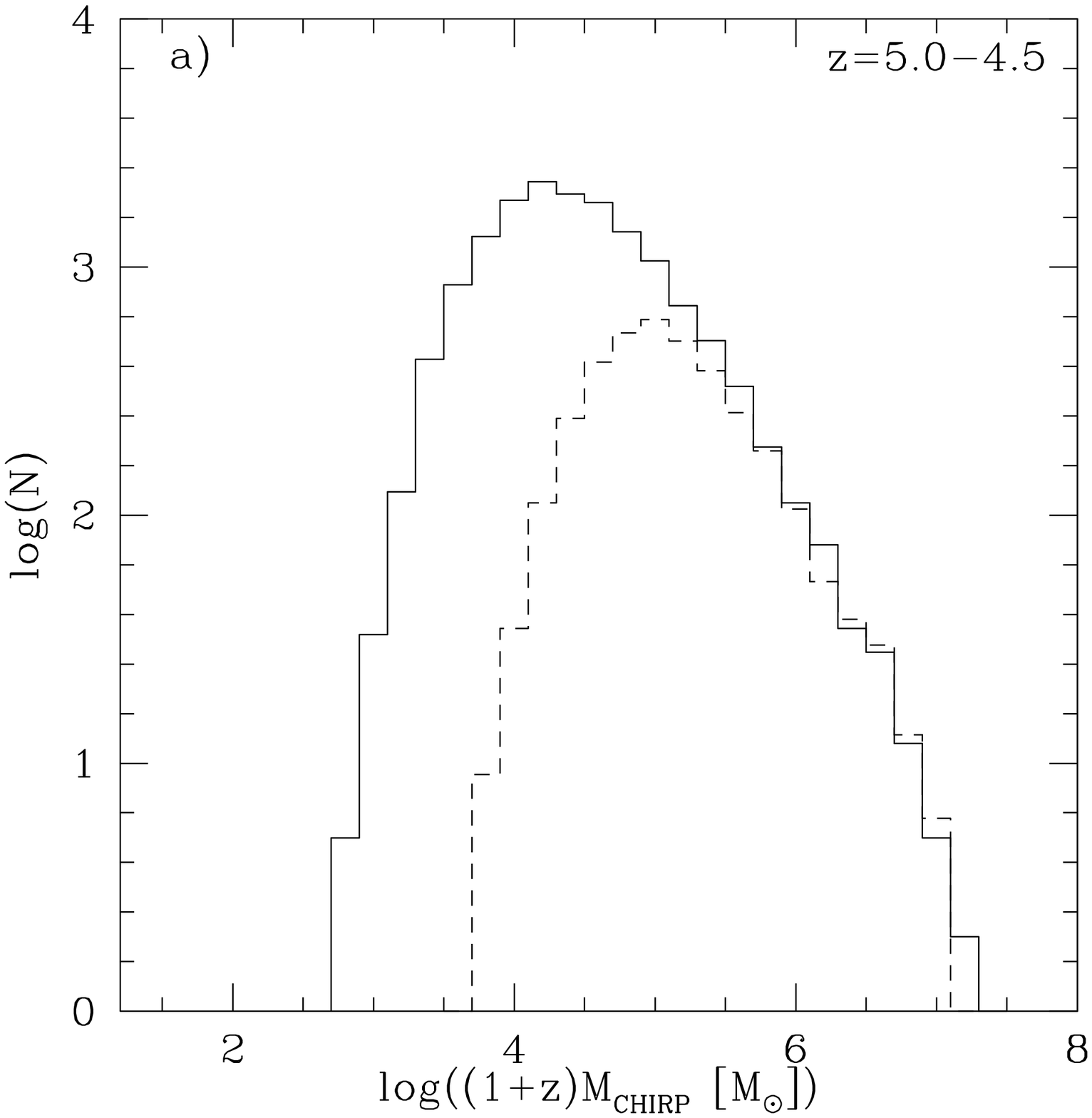,height=2.3in}
\end{displaymath}
\end{minipage}
\begin{minipage}[t]{0.45\hsize}
\begin{displaymath}
\psfig{figure=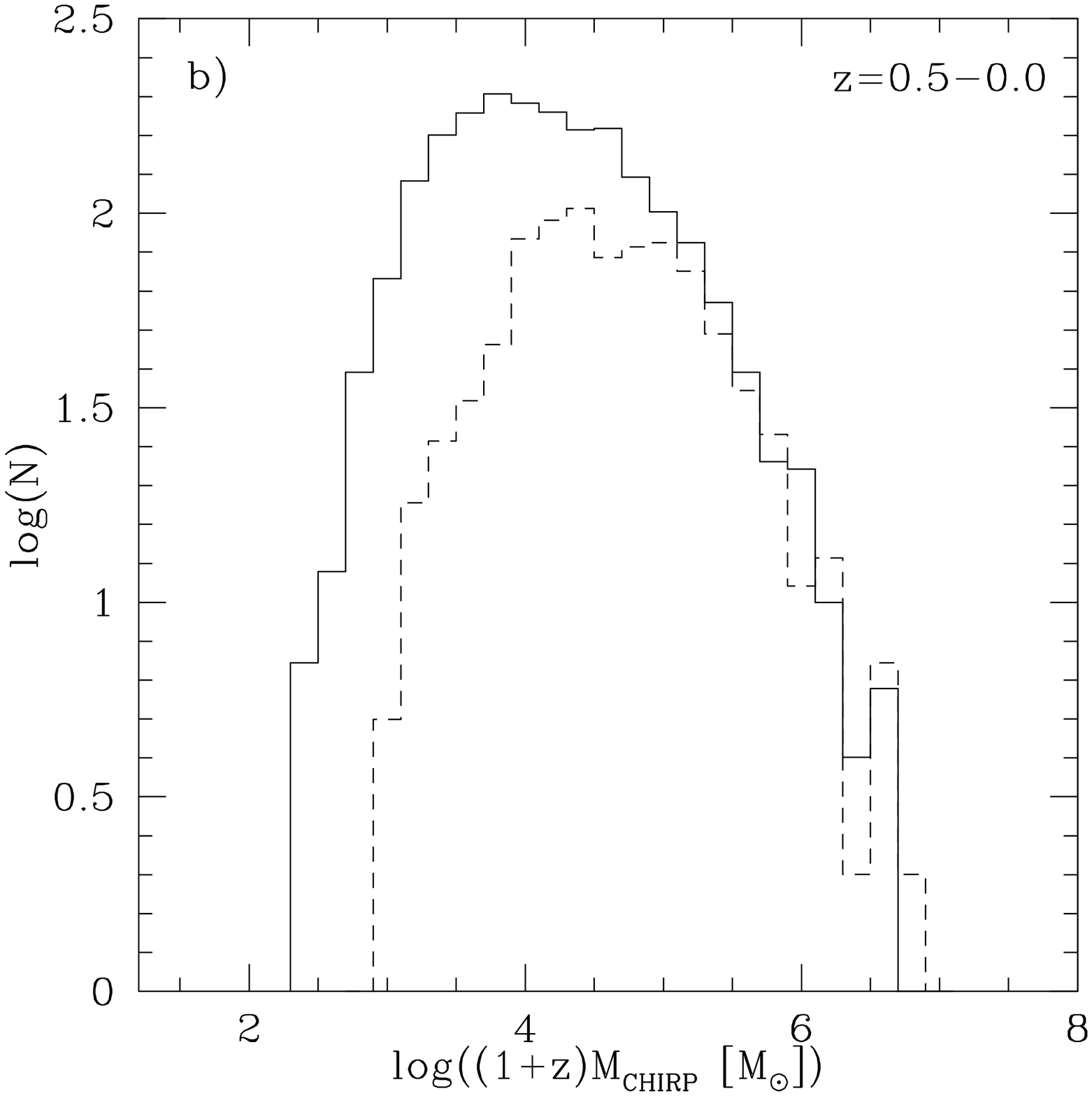,height=2.3in}
\end{displaymath}
\end{minipage}
\end{center}
\caption{\label{fig:four}Redshifted chirp mass distributions of
merging BHs in the redshift range $z=5$--$4.5$ (a) and $z=0.5$--$0$
(b), for the models with BHs in 100\% (solid) and only the 30\% most
massive (dashed) potential host galaxies at $z=5$. Normalization is
for a fixed comoving volume of $\sim 1.7 \times 10^4$~Mpc$^3$.}
\end{figure}

Although forcing the masses of MBHs to systematically follow
Eq.~(\ref{eq:one}) is arbitrary, it is partially justified by recent
results indicating that this relation may already be in place by $z
\sim 3$, at least at the high-mass end (Shields et al. 2002).
Fig.~\ref{fig:three}a and~\ref{fig:three}b show the resulting mass
distributions for merging BHs in the two models of interest, for two
representative redshift windows. The larger number of events ($\times
4$ at $z=5$, $\times 2$ at $z=0$) and the broader mass spectrum in
Fig.~\ref{fig:three}a (100\% BHs) are evident, as compared to the
model with BHs in only the 30\% most massive galaxies at $z=5$
(Fig.~\ref{fig:three}b).

Hughes (2002a) discusses the precision with which mass and redshift
measurements can be achieved with LISA, for equal-mass BH binary
mergers. Precisions of $\lsim 30 \%$ can be reached for at least two
of the three redshifted mass combinations (chirp, reduced and total
mass) in the approximate BH mass range $10^4 M_\odot/(1+z)$--$10^6
M_\odot/(1+z)$, at redshifts $z \lsim 10$ (with errors $\lsim 30\%$ on
the redshifts). Provided enough events are detected by LISA, it should
therefore be possible to distinguish between the two models shown in
Fig.~\ref{fig:three} without difficulty.

An even more exquisite precision can be achieved if one is willing to
give up the distance/redshift information and focus on redshifted
chirp masses ($M_{\rm chirp} =(m_1 m_2)^{3/5}/(m_1+m_2)^{1/5}$): those
can be determined with $1\%$ accuracy or better for equal-mass BH
binaries in the mass range $10^3$--$10^5~M_\odot$, at $z \lsim 10$.
Fig.~\ref{fig:four} compares the distributions of redshifted chirp
masses for the two models of interest, in the same two redshift
windows as before. The distribution integrated over all redshifts is
what LISA will be sensitive to, but Fig.~\ref{fig:four} nicely
illustrates how the distributions for the two models differ
significantly. The tendency for large chirp mass values in the model
with BHs in the 30\% most massive potential host galaxies at $z=5$ is
clearly seen (the same is true for the distribution integrated over
all redshifts). Provided LISA sees enough such events, it should be
very easy to distinguish between the two models (based on redshifted
chirp masses only).

\begin{figure}
\begin{center}
\hspace{-3cm}
\begin{minipage}[t]{0.45\hsize}
\begin{displaymath}
\psfig{figure=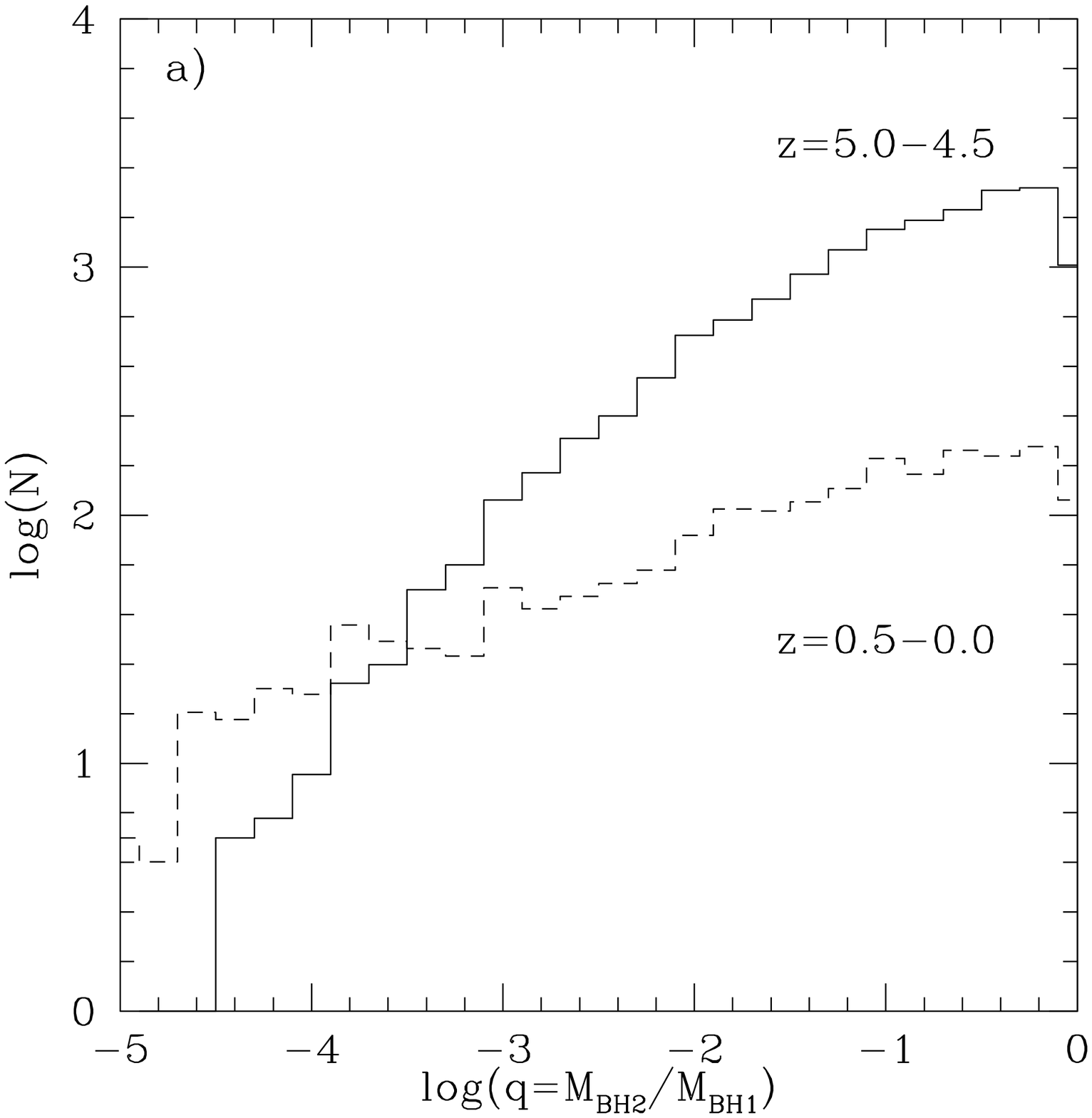,height=2.3in}
\end{displaymath}
\end{minipage}
\begin{minipage}[t]{0.45\hsize}
\begin{displaymath}
\psfig{figure=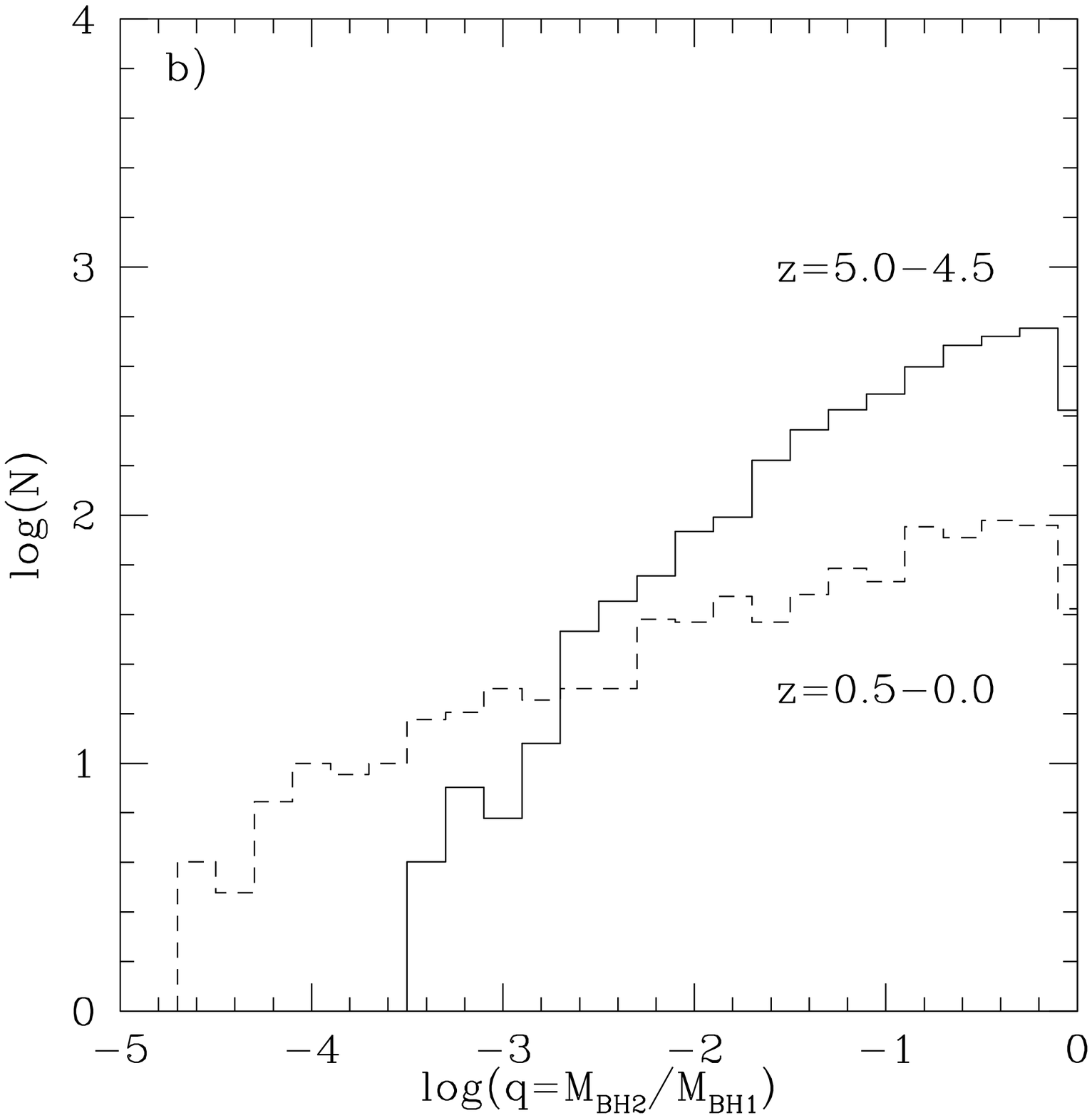,height=2.3in}
\end{displaymath}
\end{minipage}
\end{center}
\caption{\label{fig:five}Mass ratio distributions for merging BH
binaries in the redshift range $z=5$--$4.5$ (solid) and $z=0.5$--$0$
(dashed), for the models with BHs in 100\% (a) and only the 30\% most
massive (b) potential host galaxies at $z=5$. Normalization is for a
fixed comoving volume of $\sim 1.7 \times 10^4$~Mpc$^3$. }
\end{figure}

These optimistic statements ignore the following complication: the
quoted measurement errors strictly apply to equal-mass binaries
(Hughes 2002a). As Fig.~\ref{fig:five} shows, however, in both models,
the majority of MBH binary mergers are of the unequal mass type (as
expected in general). LISA measurement accuracies for equal-mass
binaries are very encouraging in suggesting that LISA will be able to
distinguish between various MBH population scenarios with exquisite
precision. Preliminary calculations for unequal mass binaries (Hughes
2002b) are promising since they suggest that precisions comparable to
those of the equal-mass case could still be achieved for (redshifted)
chirp and reduced masses.

\section{Characteristics of the MBH population probed by LISA}

There are two noticeable characteristics of the population of merging
MBHs according to the models presented in \S3 and \S4.  First,
Fig.~\ref{fig:three} shows that the mass function of merging BHs rises
steeply towards low masses. Combined with the LISA sensitivity window
peaking around $10^5 M_\odot / (1+z)$, it implies that the large
majority of the events seen by LISA should be mergers of BHs with
masses $< 10^6 M_\odot$. This contrasts with the generally larger
masses ($\gsim 10^6 M_\odot$) inferred for MBHs in bright quasars and
at the center of dynamically-studied nearby galactic nuclei. Although
the population of MBHs probed by LISA would then not dominate the mass
density of MBHs in the Universe (Yu \& Tremaine 2002), it would better
represent the total population in terms of its number density.

Second, this population of lower mass MBHs may allow us to probe some
of the properties of their low-mass host galaxies. The cutoff at low
masses in the mass function of merging BHs (see Fig.~\ref{fig:three}
and~\ref{fig:four}) reflects the assumption in the models that only
galaxies above a certain (redshift-dependent) mass threshold can host
a MBH. Indeed, studies of baryon cooling in nascent, metal-free
(primordial) galaxies suggest that, in the absence of molecular
hydrogen cooling (which is easily dissociated by a weak UV
background), the gas must rely on atomic lines to cool within a Hubble
time (e.g. Haiman et al. 2000; Loeb \& Barkana 2001). The mass (baryon
+ dark matter) of galaxies forming stars (and ``seeds'' for the MBHs,
presumably) must then be in excess of an equivalent virial temperature
$\sim 10^4$~K:
\begin{equation}
M_{\min}(z) \simeq 9 \times 10^7 M_\odot \left(\frac{T_{\rm
vir}}{10^4~K}\right)^{3/2} \left(\frac{1+z}{10}\right)^{-3/2}.
\end{equation}

Once combined with the assumption that MBHs follow the mass - stellar
(and dark matter halo) velocity dispersion relation given by
Eq.~(\ref{eq:one}) without scatter, this property translates into a
minimum BH mass in the models of
\begin{equation}
M_{{\rm BH}, \min} \sim 3000 M_\odot \left(\frac{T_{\rm
vir}}{10^4~K}\right)^{2},
\end{equation}
nearly independent of redshift. Note that this limit involves an
extrapolation of Eq.~(\ref{eq:one}) to galaxies of much lower mass
than those for which it has been observationally established. 

Hughes (2002a) shows that, for equal-mass binaries, the redshifted
chirp mass of a binary made of two $10^3~(10^4)~M_\odot$ BHs can be
determined out to $z \sim 10$ with a precision of $0.03\%~(0.07\%)$ or
better. This suggests that the location of the mass cutoff in
Fig.~\ref{fig:four} could be determined with high accuracy by LISA
(even though it will be sensitive to the distribution integrated over
all redshifts). Given the significant uncertainties in the physics of
baryon cooling (and the relevance of other effects such as UV
photo-evaporation and supernova blow-ups), the LISA sensitivity to the
BH mass cutoff could potentially be turned into a test of low-mass
galaxy cooling and formation models. Note also that arguments against
the existence of a large population of MBHs with masses $\lsim 10^6
M_\odot$ have been presented by Haehnelt et al. (1998) and Haiman et
al. (1999). LISA should be able to efficiently test these claims as
well.

\ack

The author is grateful to Pete Bender, Zolt\'an Haiman and Scott
Hughes for comments on the manuscript. Support for this work was
provided by the Celerity Foundation.

\section*{References}
\begin{harvard}
\item[] Barger A J, Cowie L L, Mushotzky R F and Richards E A 2001
{\it Astron. J.} {\bf 121} 662
\item[] Begelman M C, Blandford R D and Rees M J 1980 {\it Nature}
 {\bf 287} 307
\item[] Binggeli B, Sandage A and Tammann G A 1988 {\it
 Ann. Rev. Astron. Astrop.} {\bf 26} 509
\item[] Eisenstein D J and Loeb A 1995 {\it Astrop. J.} {\bf 443} 11
\item[] Ferrarese L and Merritt D 2000 {\it Astrop. J.} {\bf 539} L9
\item[] Filippenko A V and Sargent W L W 1989 {\it Astrop. J.} {\bf 342}
L11
\item[] Gebhardt K et al. 2000 {\it Astrop. J.} {\bf 539} L13
\item[] Gebhardt K et al. 2001 {\it Astron. J.} {\bf 122} 2469
\item[] Haehnelt M G, Natarajan P and Rees M J 1998 {\it
Mon. Not. Roy. Astron. Soc.} {\bf 300} 817
\item[] Haiman Z, Abel T and Rees M J 2000 {\it Astrop. J.} {\bf 534} 11
\item[] Haiman Z, Madau P and Loeb A 1999 {\it Astrop. J.} {\bf 514}
535
\item[] Hughes S A 2002a {\it Mon. Not. Roy. Astron. Soc.} {\bf 331} 805
\item[] Hughes S A 2002b Private communication
\item[] Kormendy J and Richstone D 1995 {\it Ann. Rev. Astron. Astrop.}
{\bf 33} 581
\item[] Loeb A and Barkana R 2001 {\it Ann. Rev. Astron. Astrop.}  {\bf
39} 19
\item[] Magorrian J et al. 1998 {\it Astron. J.} {\bf 115} 2285
\item[] Menou K, Haiman Z and Narayanan V K 2000 {\it Astrop. J.} {\bf
558} 535
\item[] Merritt D, Ferrarese L and Joseph C L 2001 {\it Science} {\bf
293} 1116
\item[] Milosavljevic M and Merritt D 2001 {\it Astrop. J.} {\bf 563}
34
\item[] Mushotzky R F, Cowie L L, Barger A J and Arnaud K A 2000 {\it
Nature} {\bf 404} 459
\item[] Quinlan G D and Hernquist L 1997 {\it New Astron.} {\bf 2} 533
\item[] Richstone D et al. 1998 {\it Nature} {\bf 395A} 14
\item[] Saslaw W C, Valtonen M J and Aarseth S J 1974 {\it Astrop. J.}
 {\bf 190} 253
\item[] Shields G A et al. 2002 {\it Astrop. J.} in press (astro-ph/0210050)
\item[] Tremaine S et al. 2002 {\it Astrop. J.} {\bf 574} 740
\item[] Volonteri M, Haardt F and Madau P 2002 {\it Astrop. J.} in
press (astro-ph/0207276)
\item[] Yu, Q 2002 {\it Mon. Not. Roy. Astron. Soc.} {\bf 331} 935
\item[] Yu, Q and Tremaine S 2002 {\it Mon. Not. Roy. Astron. Soc.}
{\bf 335} 965
\end{harvard}

\end{document}